\documentclass[%
aip,%
amsmath,amssymb,floatfix,
reprint,%
longbibliography,
]{revtex4-1}

\usepackage{graphicx}
\usepackage{dcolumn}
\usepackage{bm} 
\usepackage[usenames,dvipsnames]{xcolor}
\usepackage{soul}

\usepackage[left= 1.32cm, right = 1.32cm, top= 2.cm, bottom = 2.5cm]{geometry}

\usepackage[english]{babel} 

\usepackage{amsmath,amsthm,latexsym,amssymb,amsfonts,epsfig}
\usepackage{mathtools}
\usepackage{mathrsfs}

\usepackage{xcolor}
\definecolor{OliveGreen}{rgb}{0,0.6,0}

\newcommand{\vect}[1]{\boldsymbol{#1}}

\newcommand{\R}{\vect{r}}

\newcommand{\Intd}{\mathrm{d}}

\newcommand{\sgn}{\mathrm{sgn}}

\def\XXint#1#2#3{{\setbox0=\hbox{$#1{#2#3}{\int}$}
     \vcenter{\hbox{$#2#3$}}\kern-.5\wd0}}

\newcommand{\mhat}{\vect{\hat{m}}}
\newcommand{\rhatij}{\vect{\hat{r}}_{ij}}

\newcommand{\bOmega}{\boldsymbol{\Omega}}

\newcommand{\eX}{\vect{\hat{e}}_x}
\newcommand{\eY}{\vect{\hat{e}}_y}
\newcommand{\eZ}{\vect{\hat{e}}_z}

\newcommand{\bT}{\mu}
\newcommand{\bR}{\gamma}

\newcommand{\bP}{\mu_\mathrm{P}}

\newcommand{\sumOneQ}{\sum_{q\ge 1}}

\newcommand{\rC}{r_\mathrm{C}}

\newcommand{\yP}{y_\mathrm{P}}
\newcommand{\yPt}{y_{\mathrm{P},t}}
\newcommand{\hrho}{\hat{\rho}}
\newcommand{\hphi}{\hat{\phi}}

\usepackage[colorlinks=true,citecolor=OliveGreen,linkcolor=blue,urlcolor=blue]{hyperref}

\usepackage{epstopdf}

\usepackage{csquotes}
\usepackage{setspace}

\usepackage{type1ec} %

\allowdisplaybreaks

\begin{document}


\title{Membrane penetration and trapping of an active particle}

\thanks{Article contributed to the Topical Issue of the Journal of Chemical Physics entitled \enquote{Chemical Physics of Active Matter} edited  by Olivier Dauchot and Hartmut L\"{o}wen.}

\author{Abdallah Daddi-Moussa-Ider} 
\email{abdallah.daddi.moussa.ider@uni-duesseldorf.de}
\affiliation
{Institut f\"{u}r Theoretische Physik II: Weiche Materie, Heinrich-Heine-Universit\"{a}t D\"{u}sseldorf, Universit\"{a}tsstra\ss e 1, 40225 D\"{u}sseldorf, Germany}

\author{Segun Goh}
\affiliation
{Institut f\"{u}r Theoretische Physik II: Weiche Materie, Heinrich-Heine-Universit\"{a}t D\"{u}sseldorf, Universit\"{a}tsstra\ss e 1, 40225 D\"{u}sseldorf, Germany}

\author{Benno Liebchen}
\affiliation
{Institut f\"{u}r Theoretische Physik II: Weiche Materie, Heinrich-Heine-Universit\"{a}t D\"{u}sseldorf, Universit\"{a}tsstra\ss e 1, 40225 D\"{u}sseldorf, Germany}

\author{Christian Hoell}
\affiliation
{Institut f\"{u}r Theoretische Physik II: Weiche Materie, Heinrich-Heine-Universit\"{a}t D\"{u}sseldorf, Universit\"{a}tsstra\ss e 1, 40225 D\"{u}sseldorf, Germany}

\author{Arnold J. T. M. Mathijssen}

\affiliation{Department  of  Bioengineering,  Stanford  University,
443  Via  Ortega,  Stanford,  CA  94305,  USA}

\author{Francisca Guzm\'{a}n-Lastra}
\affiliation
{Institut f\"{u}r Theoretische Physik II: Weiche Materie, Heinrich-Heine-Universit\"{a}t D\"{u}sseldorf, Universit\"{a}tsstra\ss e 1, 40225 D\"{u}sseldorf, Germany}
\affiliation
{Facultad  de  Ciencias,  Universidad  Mayor,  Av.\  Manuel  Montt  367,  Providencia,  Santiago de Chile, Chile}

\author{Christian Scholz}
\affiliation
{Institut f\"{u}r Theoretische Physik II: Weiche Materie, Heinrich-Heine-Universit\"{a}t D\"{u}sseldorf, Universit\"{a}tsstra\ss e 1, 40225 D\"{u}sseldorf, Germany}

\author{Andreas M. Menzel}
\affiliation	
{Institut f\"{u}r Theoretische Physik II: Weiche Materie, Heinrich-Heine-Universit\"{a}t D\"{u}sseldorf, Universit\"{a}tsstra\ss e 1, 40225 D\"{u}sseldorf, Germany}

\author{Hartmut L\"{o}wen}
\email{hartmut.loewen@uni-duesseldorf.de}
\affiliation
{Institut f\"{u}r Theoretische Physik II: Weiche Materie, Heinrich-Heine-Universit\"{a}t D\"{u}sseldorf, Universit\"{a}tsstra\ss e 1, 40225 D\"{u}sseldorf, Germany}

\date{\today}

\begin{abstract}
   	
The interaction between nano- or micro-sized particles and cell membranes is of crucial importance in many biological and biomedical applications such as drug and gene delivery to cells and tissues.
During their cellular uptake, the particles can pass through cell membranes via passive endocytosis or by active penetration to reach a target cellular compartment or organelle.
In this manuscript, we develop a simple model to describe the interaction of a self-driven spherical particle (moving through an effective constant active force) with a minimal membrane system, allowing for both penetration and trapping. 
We numerically calculate the state diagram of this system, the membrane shape, and its dynamics. 
In this context, we show that the active particle may either get trapped near the membrane or penetrates through it, where the membrane can either be permanently destroyed or recover its initial shape by self-healing.
Additionally, we systematically derive a continuum description allowing to accurately predict most of our results analytically. This analytical theory helps identifying the 
generic aspects of our model, suggesting that most of its ingredients should apply to a broad range of membranes, from simple model systems composed of magnetic microparticles 
to lipid bilayers. 
Our results might be useful to predict mechanical properties of synthetic minimal membranes. 
\end{abstract}

\maketitle

\section{Introduction}

Biological membranes play a crucial role in a large variety of cellular processes, and serve as a barrier to protect the interior of living cells from unwanted agents and harmful external influences~\cite{Kaljot1988, Lodish1995, Seisenberger2001, Saar2005, Karp2008,Li2013, liboff16}. 
The interaction between particles and cell membranes is of crucial importance in a variety of biomedical applications, including 
targeted phototherapy, intracellular imaging, and diagnostic assays~\cite{xia08, kirui10, suk16}.
Once injected into a living organism, particle uptake can be achieved via passive mechanisms~\cite{lanvers17, Jirage1997, Yang2006, Kalra2003,Nelson2004} or can be mediated by active processes involving cellular energy input~\cite{nikaido92, Palacin1998, Kandel2000, Simpson2007}.
Considerable research advances have been made over the last few years in understanding the penetration of particles into cell membranes.
Previous studies have shown that the particle uptake by living cells is strongly affected by the particle properties~\cite{yu18natcomm, mathijssen2018universal, grafe16, schlenk17, muller18} and the physicochemical and functional properties of the membrane~\cite{chithrani06, lin10, yang10, dos11, dasgupta14, dasgupta17}.

As a simple framework for studying basic mechanisms of cell penetration, artificial model membranes provide a basis for understanding complex interactions within living cells.
For example, the formation of a desired target membrane structure can be driven by an entropic mechanism~\cite{Barry2010} or can be achieved using controlled external fields~\cite{zahn99, mittal09, Osterman2009, ouguz12, Dobnikar2013, williams16}.
In particular, self-assembled colloidal membranes have offered a novel framework for studying fundamental physical problems, such as geometric frustration in artificial spin-ice systems~\cite{Ortiz-Ambriz2016, Loehr2016, Tierno2016}, 
and can conveniently be built from isolated microparticles with adjustable interactions~\cite{shenton99, grzelczak10, froltsov03, froltsov05, lin05, heinrich15, yener16, peroukidis16, bharti16}.
For this purpose, various types of interparticle interactions could be exploited, among which magnetic attraction stands out.

One possibility to construct such membranes are colloidal magnetic particles, which serve as building blocks of magnetically self-assembled chains and sheets.
Magnetic nanoparticles (MNPs)~\cite{pankhurst03, lu07} are well-established nanocomponents, owing to their diverse promising technological and biomedical applications.
Notable examples include their potential use as drug delivery agents~\cite{naahidi13, al-obaidi15, liu16}, or as mediators to convert electromagnetic energy into heat (hyperthermia)~\cite{Gao2009}.
By binding MNPs to the surface of living cells, the membrane mechanical properties can conveniently be tuned by an external magnetic field~\cite{Wang1993, Dobson2008, Pankhurst2009}.
Further, magnetic colloidal and nanoparticles have proved to be useful in the design of optical stimuli-responsive materials~\cite{Ewerlin2013, Spiteri2017, Messina2014, Messina2017}, and in the development of artificial self-propelling active microswimmers~\cite{Guzman-Lastra2016, Martinez-Pedrero2015, kaiser15, babel16, mathijssen2015hydrodynamics, elgeti15, bechinger16, degraaf2016lattice, Martinez-Pedrero2018, daddi18, daddi18jpcm, Garcia-Torres2018, yu18, mathijssen18, daddi18nematic}.
Meanwhile, the dynamical properties of self-propelled active polymers and filaments have been investigated~\cite{kaiser15swell, winkler17, martin18, duman18}.
Additional works include the dynamics of semi-flexible polymer chains in the presence of nanoparticles~\cite{peng18}, and the behavior of polymers in a crowded solution of active particles~\cite{shin15}.

Here, we develop a minimal model for a (non-fluctuating) membrane made of dipolar (e.g., electric or magnetic) particles sterically interacting with a constantly driven \enquote{active} particle~\cite{tenHagen11, wittkowski12, kaiser12, wensink12, kummel13, tenhagen14, tenHagen15, speck16, driscoll18}. 
This particle may represent, e.g., a swimming microorganism~\cite{elgeti15, bechinger16} or a synthetic micro- or nanomachine that can be manipulated under the action of controlled external fields~\cite{gao13, gao14, scholz18, scholz18delay}. 
Here, we focus on the case in which the persistence length of the trajectory of the active particle is large compared to its initial distance from the membrane, i.e., the particle essentially moves along a straight line towards the membrane.

In general, active particles can reach normally inaccessible areas inside living organisms and can perform delicate and precise tasks, holding 
great promise for prospective biomedical applications such as precision nanosurgery~\cite{nelson2010, xi2013, abdelmohsen2014}, or transport of therapeutic substances to tumor and inflammation sites~\cite{wang12, wang13, paxton04}.
Direct experimental observations have recently demonstrated the self-driven motion of acoustically-powered active nanorods inside living HeLa cells~\cite{wang14acoustic}.
These nanomotors have been shown to bump into cell organelles and exhibit directional motion and spinning inside the cells.
A detailed modeling of the interactions of active particles and (cell) membranes 
may help to shed light on our understanding of the processes driving particle motion in living and synthetic cell components.
Additionally, a fundamental understanding of these processes helps to improve the controllability of micro- and nanoparticle-based 
agents in complex environments. Potentially, this might be relevant for novel therapeutic drug targets for health therapy.
One step in this direction has been taken recently specifically for self-propelled particles interacting with a moving potential interpreted as a semipermeable membrane~\cite{marini17}, 
identifying an enhanced particle accumulation in front of the
membrane accompanied with an increased drag force.
Experimentally, the mechanical pressure exerted by a set of both passive isotropic and self-propelled polar disks onto flexible unidimensional model membranes has been studied~\cite{junot17}.

In the present work, we investigate a membrane model self-assembled from dipolar spheres arranged along a chain in the two-dimensional space. 
Their dipole moment can either arise from an unscreened  magnetic or electric moment, or from screened short-ranged electric interactions, also arising from polar colloidal clusters~\cite{demirors15}.
It has previously been shown that a chain of magnetic particles can exhibit intrinsic mechanical properties reminiscent of elastic strings or rods~\cite{Vella2013, Hall2013, Kiani2015, kiani15, boltz17, deissenbeck18} depending on the additional particle interactions.
In colloidal suspensions, magnetic interactions often cause flocculation due to the strong attraction at short distances~\cite{Philipse1994}.
Such effects are usually counterbalanced by repulsive steric interactions that prevent overlapping particle volumes at finite concentrations~\cite{gast86, Nagele1996, eshraghi18}. 
Additional elastic interactions may be considered in the form of harmonic springs.
Particle systems subject to combinations of magnetic, steric, and elastic interactions have widely been utilized as a model system for ferrofluids and ferrogels~\cite{Filipcsei2007, Frickel2011, Ilg2013, Cremer2015, cremer16, Pessot2016, yannopapas16, Cremer2017, Goh2018, Menzel2018}.

Using our simple model membrane as a basis to study the penetration process by a self-driven particle (moving under the action of a constant driving force), we obtain dynamical state diagrams indicating trapping and penetration states.
We further observe penetration events with or without subsequent healing of the membrane depending on the range of the interactions between the membrane particles.
Considering a chain of dipolar spheres, we derive a continuum theory~\cite{Doi1986, Goh2018} and we probe the particle displacement and dipole reorientation caused by the self-driven particle in the small-deformation regime.
Good quantitative agreement is found between the theoretical results and numerical simulations.

The remaining part of the paper is organized as follows.
In Sec.~\ref{sec:systemSetup}, we present the system setup and derive from the potential energy the governing equations for the displacement and orientation fields of the dipolar spheres.
We then present in Sec.~\ref{sec:stateDiagram} state diagrams indicating the possible steady configurations of the system.
Moreover, we probe the transition between the dynamical states.
In Sec.~\ref{sec:analytischeTheorie}, we devise a linearized analytical theory that describes the temporal evolution of the membrane, and we provide solutions for the trapping state in Sec.~\ref{sec:AnalyticalSolutionsTrappingState}.
Concluding remarks are contained in Sec.~\ref{sec:conclusions}.

\section{System setup}
\label{sec:systemSetup}

\begin{figure}
	\begin{center}
		\includegraphics[scale=0.55]{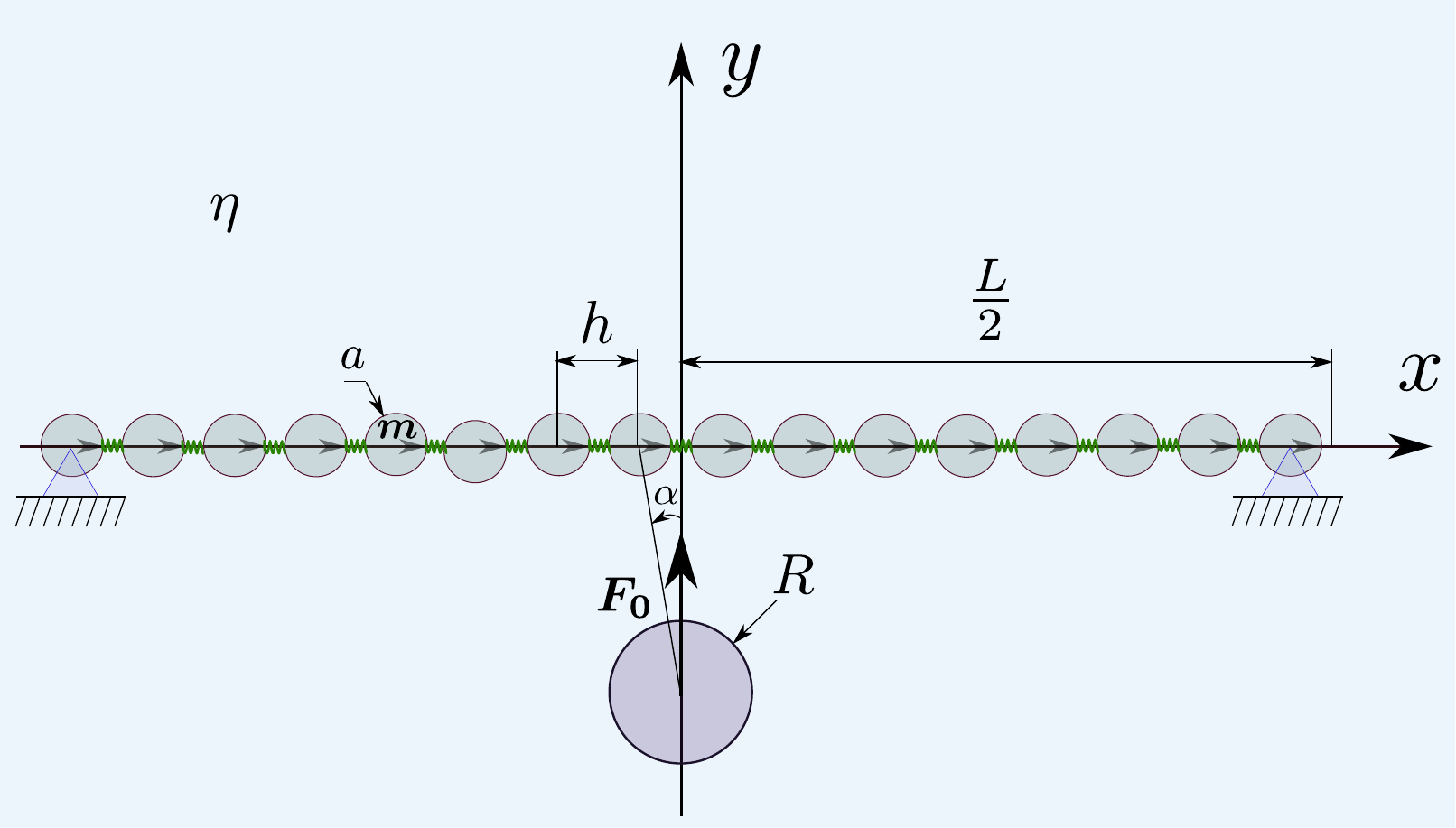}
		\caption{Illustration of the system setup.
		Under the action of an effective propulsion force~$\vect{F}_0$, a solid spherical particle of radius~$R$ approaches a membrane composed of $N$ identical magnetic spheres of radius~$a$ and dipole moment~$\vect{m}$.
		The membrane particles are initially equidistant with distance~$h$ from one another.
		We denote by~$L$ the total length of the membrane.
		The particles composing the membrane are subject to dipolar, steric, and elastic interactions.
		The system is immersed in a bulk liquid of constant dynamic viscosity~$\eta$.
		}
		\label{Sketch}
	\end{center}
\end{figure}

We consider in two spatial dimensions a simple model membrane composed of a chain of~$N$ identical dipolar particles of radius~$a$ and dipole moment~$\vect{m}$.
Here, we assume that the dipole moments rotate rigidly with the particles.
The membrane is fully immersed in a Newtonian viscous fluid of constant dynamic viscosity~$\eta$.
We support the chain at its extremities such that the particles on both ends are fixed in space.
Moreover, we neglect Brownian noise, which should play only a minor 
role when considering large membrane and large self-driven particles or systems at low temperatures.
In the resulting equilibrium configuration, the dipolar particles are uniformly distributed along the chain and aligned along the $x$~direction (Fig.~\ref{Sketch}).
We denote by $h$ the interparticle distance, initially set identical for all particles, and by~$L=hN$ the total length of the chain.

\subsection{Potential energy of the membrane}

Next, we assume that the membrane particles are subject to three types of mutual interactions, namely, dipolar, steric, and elastic interactions.
Accordingly, the system potential energy governing the time evolution of the membrane can be written as
\begin{equation}
	\mathcal{U} = \mathcal{U}_\mathrm{M} + \mathcal{U}_\mathrm{S} + \mathcal{U}_\mathrm{E} \, , 
\end{equation}
where $\mathcal{U}_\mathrm{M}$, $\mathcal{U}_\mathrm{S}$, and $\mathcal{U}_\mathrm{E}$ are contributions stemming from the dipolar, steric, and elastic interactions, respectively. 
In this study, we neglect for simplicity the fluid-mediated hydrodynamic interactions between the particles.

In the following, $\vect{m}_i$ denotes the dipole moment of the $i$th membrane particle, $i = 1, \dots, N$.
It is assumed that the magnitudes of the dipole moments are equal and constant for all the membrane particles, $m = |\vect{m}_i|$.
Then, the dipolar part of the potential energy may be expressed as~\cite{jackson12}
\begin{equation}
	\mathcal{U}_\mathrm{M} = \frac{\mu_0 m^2 }{4\pi} \sum_{\substack{i,j=1 \\ j< i}}^{N}
	\frac{1}{r_{ij}^3} \big( \mhat_i \cdot \mhat_j
	-3 \left(\mhat_i \cdot \rhatij\right) \left(\mhat_j \cdot \rhatij\right)
	 \big) \, ,
\end{equation}
where~$\mu_0$ is the magnetic vacuum permeability, $\mhat_i = \vect{m}_i/m$ gives the orientation of the $i$th dipole moment, $\vect{r}_{ij} = \R_i - \R_j$ denotes the distance vector from particle $j$ to particle $i$, $r_{ij} = |\vect{r}_{ij}|$ is its magnitude, and $\rhatij = \vect{r}_{ij}/r_{ij}$ stands for the corresponding unit vector.

In order to avoid aggregation of the dipolar particles, we consider a repulsive Weeks--Chandler--Andersen (WCA) pair potential.
The corresponding potential energy reads~\cite{weeks71}
\begin{equation}
	\mathcal{U}_\mathrm{S} = 4\epsilon \sum_{\substack{i,j=1 \\ j< i}}^{N} N_{ij}  
	\left(\frac{\sigma}{r_{ij}}\right)^{6}
	\left( \left(\frac{\sigma}{r_{ij}}\right)^{6}
	- 1 \right) + \epsilon \, , \label{hamiltonianWCA}
\end{equation}
where we have defined the shorthand notation $N_{ij} = H\left(\rC-r_{ij}\right)$, with $H(\cdot)$ being the Heaviside step function and $\rC = 2^{1/6}\sigma$ denoting a cutoff radius beyond which the potential energy is set to zero.
Here, $\sigma=2a$ is the particle diameter, and $\epsilon$ is an energy scale associated with the hardness of the potential.

In addition, we allow for harmonic elastic-like interactions among adjacent particles.
These interactions are included as springs of constant stiffness~$k$ and rest length~$r_0$. 
The corresponding potential energy is given by
\begin{equation}
	\mathcal{U}_\mathrm{E} = \frac{k}{2} \sum_{i=1}^{N-1} \left( r_{i, i+1} - r_0 \right)^2 \, .
\end{equation}

Consequently, the resulting force and torque acting on the $i$th sphere are calculated from the system potential energy as~\cite{babel16} $\vect{F}_i = -\partial \mathcal{U} / \partial \R_i$ and $\vect{T}_i = -\mhat_i \times \left(\partial \mathcal{U} / \partial \mhat_i \right)$.
We obtain
\begin{align}
			\vect{F}_i  &= \frac{3\mu_0m^2}{4\pi} \sum_{\substack{j=1 \\ j\neq i}}^{N}
					\frac{1}{r_{ij}^4} \big( 
					 \left( \mhat_j \cdot \rhatij \right) \mhat_i
					+ \left( \mhat_i \cdot \rhatij \right) \mhat_j \notag \\
					&+  \left( \mhat_i \cdot \mhat_j \right) \rhatij 
					-5 \left( \mhat_i \cdot \rhatij \right) \left( \mhat_j \cdot \rhatij \right) \rhatij
					 \big)   \notag \\
			 &+ 48\epsilon \sum_{\substack{j=1 \\ j\neq i}}^{N} N_{ij} \left(\frac{\sigma}{r_{ij}}\right)^6
					 \left( \left(\frac{\sigma}{r_{ij}}\right)^6- \frac{1}{2} \right)
					 \frac{\rhatij}{r_{ij}} \notag \\
		    &+ k \sum_{\substack{j=i-1 \\ j \ne i}}^{i+1}
		    	 \left( \frac{r_0}{r_{ij}} - 1 \right) \vect{r}_{ij} \label{ForceEq}
\end{align}
and 
\begin{equation}
		\vect{T}_i = - \frac{\mu_0 m^2}{4\pi} \sum_{\substack{j=1 \\ j\neq i}}^{N}
					\frac{\mhat_i \times \vect{c}_{ij}}{r_{ij}^3}   \, , 
						 	\label{TorqueEq}
\end{equation}
where we have defined, for convenience, the dimensionless vector $\vect{c}_{ij} = \mhat_j - 3  \left( \mhat_j \cdot \rhatij \right) \rhatij$.


\subsection{Dynamical equations}

Assuming low-Reynolds-number hydrodynamics~\cite{happel12}, the moments of the particle velocities are related to the moments of the hydrodynamic forces acting on them via the mobility functions~\cite{kim13, balboa17}.
Neglecting mutual hydrodynamic interactions between the particles yields
\begin{equation}
	\vect{V}_i = \bT \left( \vect{F}_i + \vect{F}^\mathrm{ext}_i \right) \, , \qquad 
	\bOmega_i = \bR \, \vect{T}_i \, , \label{TranslationalRotationalDef}
\end{equation}
where~$\vect{V}_i$ and $\bOmega_i$ denote the linear and angular velocities of the $i$th membrane particle, respectively.
Here, $\bT = 1/(6\pi\eta a)$ and $\bR = 1/(8\pi\eta a^3)$ are, respectively, the translational and rotational mobilities for a sphere as given by the Stokes formulas.
Moreover, $\vect{F}^\mathrm{ext}_i$ is the external force resulting from the steric interaction with the self-driven particle that is moving under the action of a constant driving force~$\vect{F}_0 = F_0 \,\eY$. 
Here, we assume that the self-driven spherical particle of radius~$R$ interacts with membrane particles via the same soft repulsive WCA pair potential stated by Eq.~\eqref{hamiltonianWCA}, for $\sigma= a+R$.

Then, the equation of motion for the translational degrees of freedom reads  
\begin{equation}
		\frac{\Intd \R_i}{\Intd t}    = \vect{V}_i \, . \label{translationalVelo}
\end{equation}
Similarly, the equation governing the temporal evolution of the orientation of the $i$th particle is given by
\begin{equation}
	\frac{\Intd \mhat_i}{\Intd t}  = \bOmega_i \times \mhat_i \, , 
\end{equation}
which can be rewritten as
\begin{equation}
	\frac{\Intd \mhat_i}{\Intd t}
	= \frac{\bR \mu_0 m^2}{4\pi}  
	\sum_{\substack{j=1 \\ j\neq i}}^{N}
	\frac{1}{r_{ij}^3}
    \big( \left(\mhat_i \cdot \vect{c}_{ij}\right) \mhat_i - \vect{c}_{ij} \big) \, , 
\end{equation} 
by making use of Eqs.~\eqref{TorqueEq} and \eqref{TranslationalRotationalDef}.

Considering now two-dimensional orientation vectors in the $(xy)$ plane, the particle orientations are represented in the Cartesian basis system as $\mhat_i = (\cos\phi_i , \sin\phi_i)$ with the angle $\phi_i$ measured relatively to the $x$ direction.
Furthermore, the angular velocity vector then possesses only one single component (along the $z$ direction).
Hence, the temporal evolution of the orientation angle of the $i$th particle is calculated as
\begin{equation}
	\frac{\Intd \phi_i}{\Intd t} = \bOmega_i \cdot \eZ =
	\bR \left( \vect{T}_i \cdot \eZ \right) \, .
	\label{torqueBalance}
\end{equation}

We now introduce an additional cutoff length $\ell = 3h/2$ for the dipolar and elastic
interactions. That is, we multiply a Heaviside function of the form $H(\ell - r_{ij})$
to Eqs.~\eqref{ForceEq} and~\eqref{TorqueEq}.  
Accordingly, these interactions are now truncated beyond next-nearest neighbors.
Such a cutoff can be reasonable for screened electric dipolar interactions.
This assumption does not significantly change our results except for the membrane destruction state of absent-healing (see below), the occurrence of which hinges on the cutoff.

\section{State diagram}
\label{sec:stateDiagram}

\begin{figure}
\begin{center}
\includegraphics{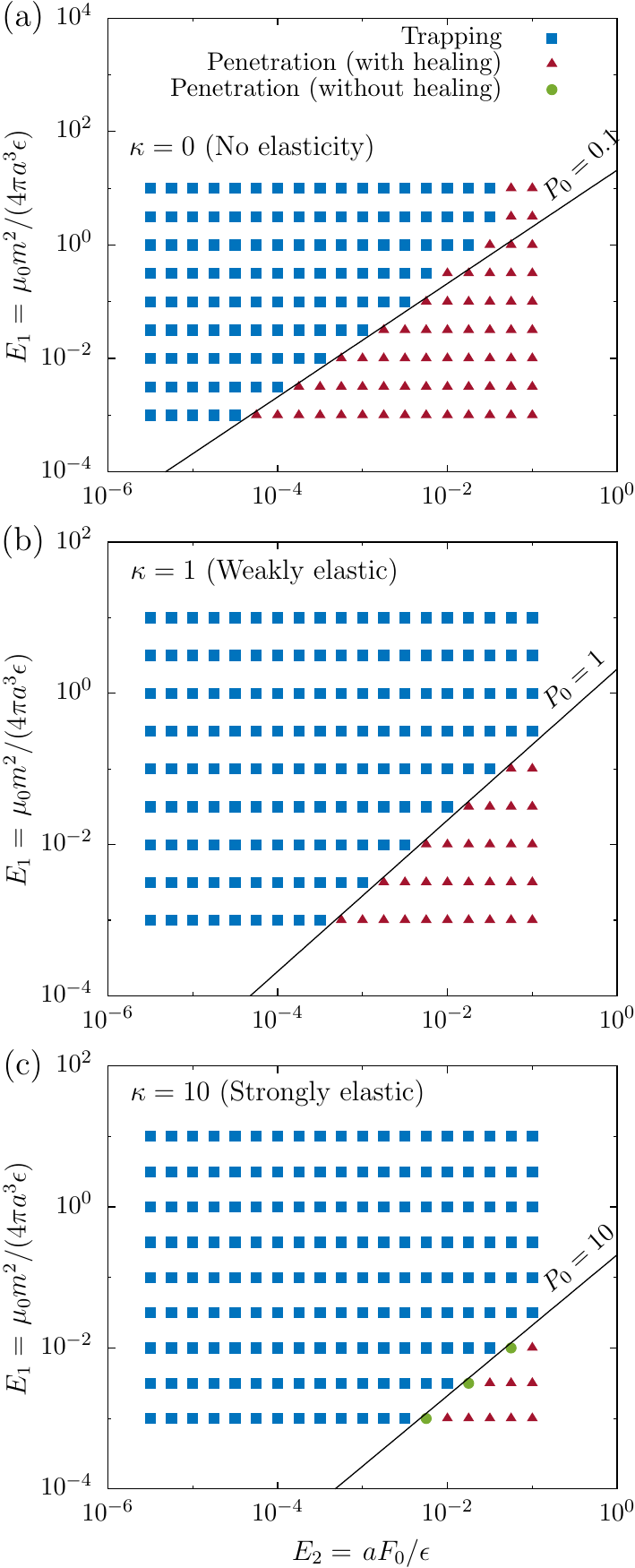}
\end{center}
\caption{(Color online) 
Ability of penetration or trapping as a function of elasticity.
Shown are state diagrams for $(a)$ $\kappa=0$, $(b)$ $\kappa=1$, and $(c)$ $\kappa=10$. 
Symbols represent the final states obtained from numerical integration of the dynamical equations, given by Eqs.~\eqref{ForceEq}--\eqref{torqueBalance}.
Here, membranes consisting of $N=20$ dipolar particles have been examined, and we set the size ratio $\delta=1$.
Depending on the values of the dimensionless numbers~$E_1$ and~$E_2$, the active particle is either trapped (blue squares), or passes through the membrane to reach the other side.
After full penetrations, the membrane either shows a self-healing ability (red triangles) or remains permanently damaged (green disks).
The latter behavior is only observed in the case of strongly elastic membranes shown in~$(c)$, for the present set of parameters.
The solid lines display estimates of the transition line between the states.
}
\label{State-Diagram-Spring}
\end{figure}

\begin{figure*}
\begin{center}
\includegraphics{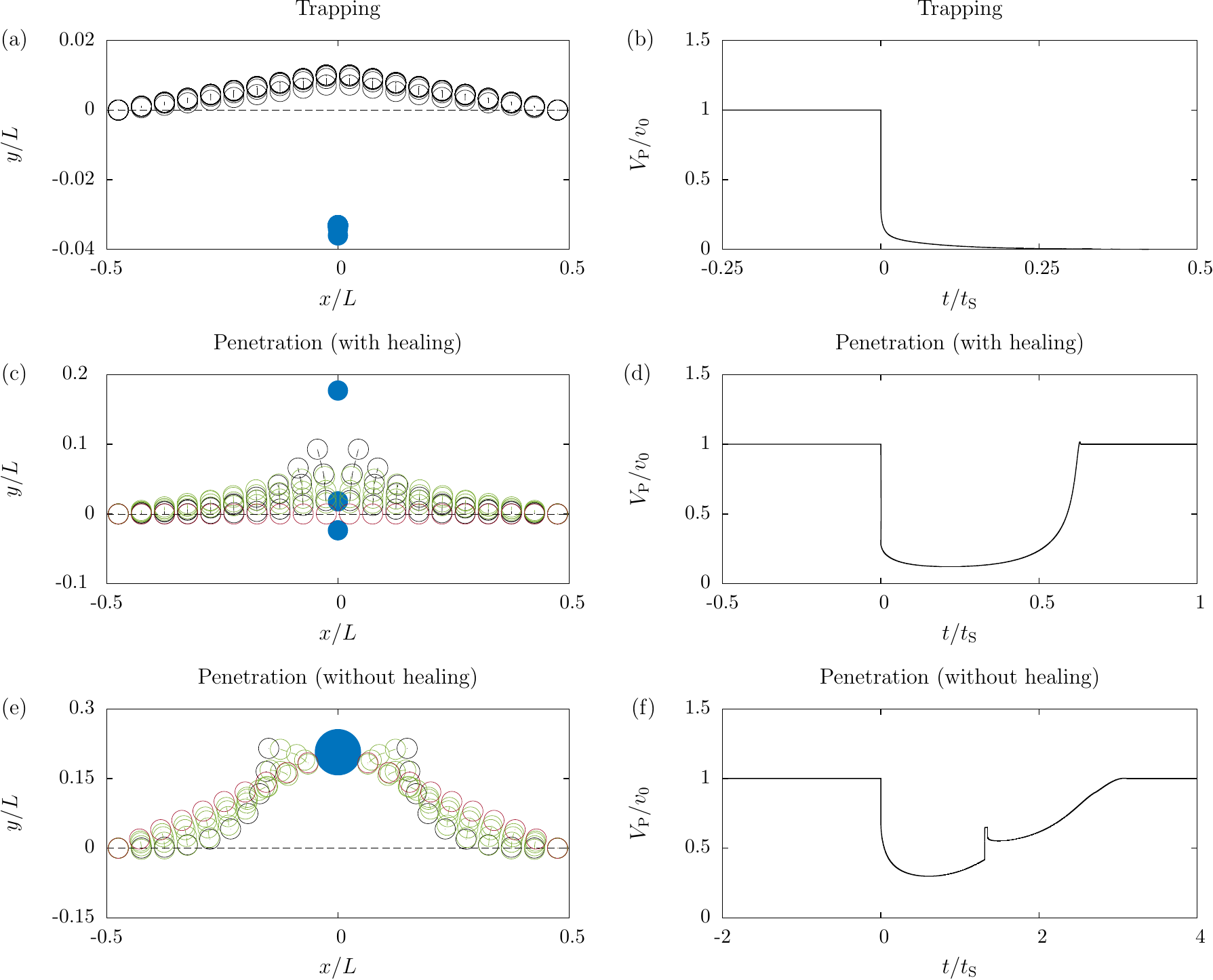}
\end{center}
\caption{(Color online)
Membrane dynamics of trapping and penetration states.
$(a)$~Frame series in the trapping state for $N=20$, $\kappa=0$, $\delta=1$, $E_1=1$, and $E_2=10^{-2}$.
Here, the frames are displayed every $0.2 \, t_\mathrm{S}$, where $t_\mathrm{S}=\eta L^3 / \epsilon$ is the simulation time unit.	
$(b)$~Time evolution of the translational velocity of the active particle in the trapping state.
$(c)$~Frame series of the membrane conformation during the penetration state with healing, using the same set of parameters as in $(a)$, except for $E_1=0.1$.
The frames are displayed in time every $0.6 \,t_\mathrm{S}$.
The black and green circles represent the positions of the membrane particles, respectively, before and after the active particle (blue disk) reaches the upper side. 
As shown, the membrane recovers its original conformation after the active particle has passed (red circles).
Panel~$(d)$ shows the corresponding translational velocity of the active particle versus time.
$(e)$~Frame series of the membrane shape during the penetration state without healing, using the same parameters as in~$(c)$, except for $R=5a$.
The frames are displayed every $6 \, t_\mathrm{S}$ in time with the same color as in $(c)$.
Circles shown in red represent the steady positions of the membrane particles.
Panel~$(f)$ displays the corresponding time evolution of the active particle.
We note that the particles in $(a)$, $(c)$, and $(e)$ are not plotted to scale.
Accordingly, the shown circles and disks only correspond to the positions of the centers of the particles.
(The membrane particles and the driven particle in~$(a)$ are actually in contact, but the scales on the ordinate and abscissa are pronouncedly different).
Time $t=0$ in the subfigures $(b)$, $(d)$, and $(f)$ corresponds to the moment when the active particle and the membrane begin to mutually interact.
}
\label{MemShape}
\end{figure*}

As an initial configuration of the membrane, the interparticle distance $h$ is taken equal to the cut-off radius $r_\mathrm{C}$ beyond which the steric forces vanishes.
Moreover, we assume that the rest length of the springs is equal to this initial interparticle equilibrium distance, i.e., $r_0 = 2^{7/6} a$.

Our parameter space has four essential dimensions.
The two dimensionless numbers
\begin{equation}\label{DefinitionE1E2}
	E_1 = \frac{\mu_0 m^2}{4\pi a^3\epsilon} \, ,  \qquad 
	E_2 = \frac{a F_0}{\epsilon}
\end{equation}
quantify, respectively, the importance of the attractive dipolar force $(\sim \mu_0 m^2/a^4)$ and of the active force $F_0$ relative to the repulsive steric force $(\sim \epsilon/a)$ at particle contact.
These two parameters will, respectively, be denominated as reduced dipole strength and reduced activity.
One additional dimensionless number
\begin{equation}
	\kappa = \frac{\pi}{6} \frac{k h^5}{\mu_0 m^2}
\end{equation}
corresponds to the ratio of the elastic to the dipolar interactions.
Moreover, we define the dimensionless number
\begin{equation}
	\delta= \frac{R}{a}
\end{equation}
as the ratio of the radius of the active particle relative to that of the membrane particle.
The parameters~$\kappa$ and $\delta$ will be denominated as reduced stiffness and size ratio, respectively.
For future reference, we also introduce a dimensionless number quantifying the ratio of the driving and dipolar forces in the form
\begin{equation}
	P_0 = \frac{1}{12} \left( \frac{h}{a} \right)^4 \frac{E_2}{E_1} \, .
	\label{P0Def}
\end{equation}
The latter will serve as our key control parameter discriminating trapped from penetrating states as detailed below.
We note that $h/a=2^{7/6}$ is kept constant such that $P_0$ is fully determined from the ratio $E_2/E_1$.

In Fig.~\ref{State-Diagram-Spring}, we present state diagrams identifying the possible dynamical states of the system in the plane of the two control parameters~$E_1$ and $E_2$.
The diagrams are constructed by numerical integration of the dynamical equations of motion using a 4th-order Runge-Kutta scheme with adaptive time step~\cite{press89}.
Results are shown for three values of the reduced stiffness~$\kappa$ which span a wide range of values to be expected in various situations.
Here, we set $N=20$ and  $\delta=1$.
We have tested the robustness of the state diagrams by varying the number of membrane particles and have found no qualitative difference.
Depending on the combination of the relevant control parameters, the self-driven particle either penetrates, or remains in direct contact with the membrane (trapping state).
In the latter case, the particle is essentially held back due to the steric interactions with the membrane particles.
Furthermore, two penetration regimes are identified depending on whether the membrane self-heals and recovers its initial undeformed shape (red triangles) or remains damaged after the particle reaches the other side (green disks in~$(c)$).
Qualitatively, penetration scenarios are observed for higher values of~$P_0$ that indicate larger driving forces or smaller restoring dipolar forces than those in the trapped state.
For $\kappa \gg 1$, penetration happens when 
\begin{align}
\label{eq:criticalForce}
\frac {P_0}{\kappa} = \frac{2F_0}{k h} \gtrsim 1 \, ,
\end{align}
i.e., when the active force is larger than the overall elastic and dipolar restoring forces of a membrane particle with its two neighbors. 
After membrane penetration, self-healing always occurs for non- or weakly-elastic membranes, for the present set of parameters.
In contrast to that, the membrane may remain permanently damaged for strongly elastic membranes, see Fig.~\ref{State-Diagram-Spring}~$(c)$.
Besides, the elastic interactions cause a noticeable `shifting' of the transition line between the penetration and trapping states.
Apart from that, they do not qualitatively alter our results and will therefore be omitted in most of our later calculations.
It is worth noting that the detailed form of the steric repulsion may not be important as long as the reduced dipole strength $E_1 \ll 1$.
An alternative could be the use of hard-core interactions.
However, a softer potential is adopted here for numerical convenience to prevent the interparticle forces from diverging during the evolution dynamics.

We now describe the dynamical scenarios of the trapped and penetrating states depicted in Fig.~\ref{MemShape}.
First, we examine the time evolution of membrane configurations.
At the initial stage of the dynamics, the active particle pushes the membrane
and subsequently bends the membrane, as can be seen in Fig.~\ref{MemShape} $(a)$, $(c)$, and $(e)$.
If the active force is strong enough ($P_0 \gg 1$), such deformation persistently increases and induces a growing distance between the two center particles of the chain, giving rise to a weakening of their mutual dipolar attraction.
Consequently, the active particle penetrates through the membrane, see Fig.~\ref{MemShape}~$(c)$ and~$(e)$.
Depending on the size of the active particle relative to that of the membrane particles, the membrane either closes again to recover its initial aligned configuration (self-healing behavior shown in $(c)$ for $\delta=1$) or remains permanently deformed (as shown for $\delta=5$ in $(e)$).
In addition, we observe that the penetration event is also accompanied by a slight abrupt increase in the particle speed (small cusp occurring in $(d)$ at $t/t_\mathrm{S} \simeq 0.6$ and in $(f)$ at $t/t_\mathrm{S} \simeq 1.6$).
This small augmentation of speed is due to the steric interactions which support the particle motion at this final stage when the penetrated particle is sterically repelled by the nearby membrane particles.
In sharp contrast, when $P_0 \ll 1$, the membrane develops a triangular profile, reaching a steady state without allowing the self-driven particle to pass, see Fig.~\ref{MemShape}~$(e)$.
This trapping behavior is investigated in more details in Secs.~\ref{sec:analytischeTheorie} and \ref{sec:AnalyticalSolutionsTrappingState}.
Meanwhile, both scenarios can also be understood in terms of the velocity profiles of the self-driven particle presented in Fig.~\ref{MemShape} $(b)$, $(d)$, and $(f)$.
Since the dynamics are overdamped, the velocity can be interpreted as the total net force exerted on the particle.
Accordingly, membrane penetration occurs when the external driving force remains larger than the membrane restoring forces.

\begin{figure}
\begin{center}
\includegraphics{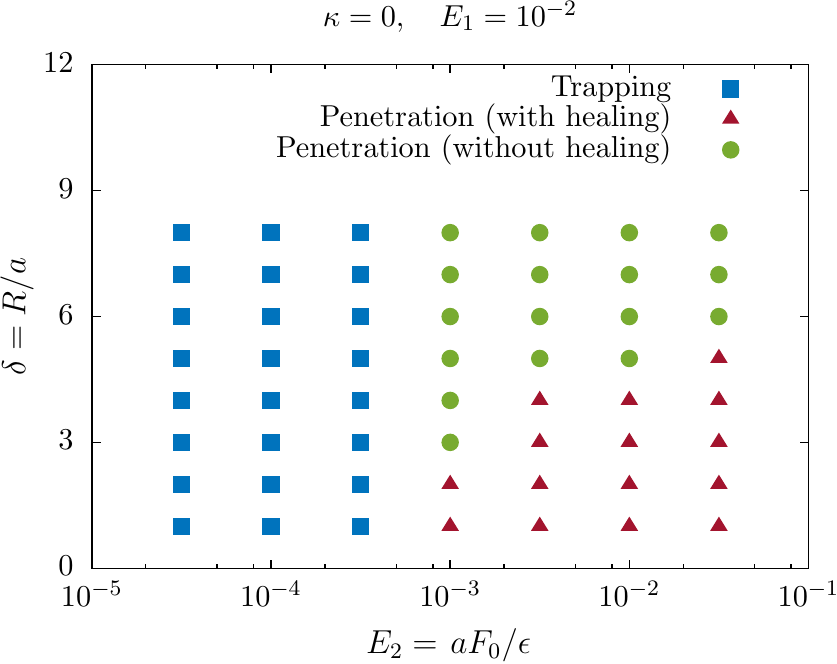}
\end{center}
\caption{(Color online) 
State diagram of trapping and penetration in the parameter space of size ratio~$\delta$ and reduced activity~$E_2$, while keeping the reduced stiffness $\kappa = 0$ and the reduced dipole strength $E_1 = 10^{-2}$.
Here, we have examined membranes consisting of $N=20$ dipolar particles.
Symbols represent the final dynamical state obtained from numerical integration of Eqs.~\eqref{ForceEq}--\eqref{torqueBalance}.  
}
\label{State-Diagram-Radius}
\end{figure}

In order to explore the membrane behavior in the penetration state in more detail, we present in Fig.~\ref{State-Diagram-Radius} a state diagram in the parameter space $(\delta, E_2)$.
Here, we keep the other parameters fixed at $\kappa=0$, $E_1=10^{-2}$, and $N=20$.
We observe that the transition between the trapping and penetration states can only be enabled by increasing the reduced activity~$E_2$, regardless of the size ratio~$\delta$.
However, the latter strongly affects the membrane behavior in the penetration scenario.
In the considered range of parameters, lower values of~$\delta$ lead to self-healing, while larger values imply permanent damage of the membrane.
The observed suppression of the healing behavior for large enough penetrating particles can be understood by the fact that the mutual distance between the two central beads becomes larger than the cutoff distance~$\ell$, which could represent the average distance between cytoskeletal cross-linkers for biological membranes.
If these links are broken by large active particles, the attractive interactions between the membrane particles vanish.
Consequently, the membrane is split up and remains permanently destroyed, or at least until other mechanisms help the membrane to regenerate.
Without the cutoff, the membrane because of the long-ranged forces always heals after a penetration event.


\section{Analytical theory}
\label{sec:analytischeTheorie}

To proceed analytically, we restrict ourselves to the small-deformation regime.
Then, we linearize the dynamical equations and solve for the membrane displacement and dipole orientation fields.

\subsection{Evolution of the membrane particles}

In the deformed configuration, the position vector of each dipolar particle in the laboratory frame of reference can be written as $\R_i=(d_i+u_i)\eX+\rho_i \eY$, 
wherein $d_i = h (i-N/2)$, $i=1, \dots, N$, represents the equilibrium $x$~positions of the particles in the initial configuration.
Without loss of generality, we consider here only even numbers of~$N$.
In addition, $u_i$ and $\rho_i$ denote the membrane displacements along the $x$ and $y$ directions, respectively.


We assume that the active particle has a radius comparable to that of the membrane particles.
For the dipolar particles that are not at the chain ends, i.e., for $i = 2, \dots, N-1 $, the projection of the dynamic equations governing the translational motion of the $i$th sphere, given by Eq.~\eqref{translationalVelo}, can be presented in a linearized form as
\begin{subequations} \label{discreteDisplacements}
	\begin{align}
		\frac{1}{A} \frac{\Intd u_i}{\Intd t} &= 
		\frac{72\epsilon\mu}{A h^2}
			\big( (u_{i+1}-u_i) N_{i,i+1} - (u_i-u_{i-1}) N_{i,i-1} \big) \notag \\
		&+2\left( \kappa-1  \right) \frac{u_{i+1}-2u_i+u_{i-1}}{h^2} - \frac{\mu  {{F}_\parallel}_i}{A} \, , 
		\label{discreteDisplacementsU} \\
		\frac{1}{A} \frac{\Intd \rho_i}{\Intd t} &= 
		 \frac{\rho_{i+1}-2\rho_i+\rho_{i-1}}{h^2}  
		- \frac{\phi_{i+1}-\phi_{i-1}}{4h}
		+ \frac{\bT {{F}_\perp}_i}{A} \, , \label{discreteDisplacementsRho}
	\end{align}
\end{subequations}
where we have defined $A:=3\mu_0 m^2 \bT/(\pi h^3)$, a parameter that has the dimension of a diffusion coefficient.
We assume that $r_{i, i \pm 1} < \ell$ always holds in the small-deformation regime considered here.
Moreover, ${F_\parallel}_i = F_i \sin\alpha$ and ${F_\perp}_i = F_i \cos\alpha$,
where $F_i = F \left( \delta_{i,N/2} + \delta_{i,N/2+1} \right)$ is the magnitude of the force acting on the two central particles due to the steric interactions with the active particle.
Thus, $F_i=F$ if $i \in \{N/2, N/2+1\}$, and $F_i=0$ otherwise. 
This implies that the active particle is exactly positioned between the central two beads of the membrane.
We have also explored the situation where $N$ is an odd number, in which the external force is only exerted to the center particle, and have found quantitatively similar results.
Continuing, $\alpha$ is the angle formed by the $y$~axis and the line connecting the center of the self-driven particle to that of the closest membrane particle (see Fig.~\ref{Sketch}). 
This angle is defined as negative for clockwise rotation from the $y$~axis.
Notably, the dipolar interactions manifest themselves in both the longitudinal and transverse force balance equations, whereas the steric and elastic interactions are (at linear order) only involved 
in the longitudinal force balance equation.
This behavior resembles that of elastic membranes, where stretching and bending effects are predominately pronounced along the tangential and normal traction jumps, respectively~\cite{daddi16, daddi17, daddi2017thesis, daddi18epje, daddi18stone}.
Therefore, our self-assembled chains can be used as a minimal model membrane with effective stretching and bending moduli, in analogy to purely elastic membranes with stretching and bending deformation modes.

Similarly, we proceed with the torque balance given by Eq.~\eqref{torqueBalance}, and derive an approximate equation for the rotational motion of the membrane particles.
Upon linearization, we obtain
\begin{equation}\label{discreteOrientation}
	\frac{\Intd \phi_i}{\Intd t} = \frac{B}{2}
	 \left( \frac{\rho_{i+1}-\rho_{i-1}}{h} - \frac{\phi_{i+1}+\phi_{i-1}+4\phi_i}{3} \right) \, ,
\end{equation}
where we have defined a parameter $B := 3A/ \left( 8a^2 \right)$ with the dimension of inverse time.
We also used the fact that the translational and rotational mobilities of a sphere are related via $\bR / \bT = 3/(4a^2)$.

The two particles located at the membrane extremities remain fixed in space (zero displacement) and not subject to any dipolar torques.
The latter could be achieved, for instance, if for the two particles at the ends of the membrane the dipole moment can freely rotate inside the particle, relatively to the particle frame.
Therefore, Eqs.~\eqref{discreteDisplacements} and \eqref{discreteOrientation} are subject to the boundary conditions 
\begin{subequations}\label{discreteRandbedingungen}
	\begin{align}
		u_i = \rho_i = 0 \, , \text{~~for~~} i \in \, & \{1,N\} \, ,   \\
		\phi_2+2\phi_1 - 3\,\frac{\rho_2}{h} &= 0 \, , \\
		\phi_{N-1} + 2\phi_N + 3\, \frac{\rho_{N-1}}{h} &= 0 \, .	
	\end{align}
\end{subequations}

\subsection{Evolution of the active particle}

The active particle is subject to the constant force~$\vect{F}_0$ acting along the $y$ direction in addition to the resistive forces due to the steric interactions with the two central particles.
Denoting by $\mu_\mathrm{P} = 1/(6\pi\eta R)$ the translational mobility function of the self-driven particle, the governing equation for the translational motion along the $y$ direction reads
\begin{equation}
	\frac{1}{\bP} \frac{\Intd \yP}{\Intd t} =  F_0 - 2 F \cos \alpha \, . \label{externeTeilchengleichung}
\end{equation}

For future reference, we define~$r$ as the steady center-to-center distance separating the self-driven particle from the central particles in the trapping state.
For an interparticle distance $r\lesssim \rC$, the magnitude of the WCA force acting on a central particle can, to leading order, be approximated by 
\begin{equation}
	F = \frac{36 \cdot 2^{2/3} \epsilon}{\sigma} \left( 2^{1/6}-\frac{r}{\sigma} \right) \, ,
\end{equation}
where $\sigma = a+R$.

Inserting the latter equation into Eq.~\eqref{externeTeilchengleichung} and setting the left-hand side to zero, the steady-state distance separating the self-driven particle from the central particles is given by
\begin{equation}
	r = \frac{h}{2} \left(1+\frac{R}{a}\right)
	\Bigg( 1 - \frac{E_2}{288}  \left(1+\frac{R}{a}\right) \frac{h}{a \cos\alpha}  \Bigg) \ ,
	\label{rFormula}
\end{equation} 
where we have used the constraint that $h/a=2^{7/6}$.

Eqs.~\eqref{discreteDisplacements} and \eqref{discreteOrientation} form $3(N-2)$ ordinary differential equations in time for the unknown displacement and orientation fields.
These equations are subject to the six boundary conditions given by Eqs.~\eqref{discreteRandbedingungen} in addition to the initial conditions of vanishing displacement and orientation fields.
In the steady state, the problem reduces to finding the solution of a set of recurrence equations relating the positions and orientations of adjacent spheres.
In Sec.~\ref{sec:AnalyticalSolutionsTrappingState}, we present an analytical solution of the resulting recurrence problem.
In addition, we show that the underlying equations for the motion of the membrane particles can conveniently be presented in the continuous limit using partial differential equations that describe the temporal and spatial evolution of the membrane displacement and dipole orientation.



\section{Solution for the trapping state}
\label{sec:AnalyticalSolutionsTrappingState}

\subsection{Steady solution of the recurrence problem}

For $E_2 \ll 1$, it follows from Eq.~\eqref{rFormula} that $r\sim h(1+\delta)/2$, where again $\delta = R/a$.
Assuming that $|u_i| \ll h$, for $i=1, \dots, N$, yields $\sin\alpha \simeq h/(2r)$.
As a result, $\alpha \simeq \arcsin \left( 1/(1+\delta) \right)$.

Due to the symmetry of the problem with respect to the membrane center, it is sufficient to solve the recurrence problem for $i \in \{1, \dots, M\}$, where $M:=N/2$.
In the steady state, it follows readily from the force balance Eq.~\eqref{externeTeilchengleichung} that $F=F_0/(2\cos\alpha)$, where $\cos\alpha \simeq \left( 1-1/(1+\delta)^2 \right)^{1/2}$.

\subsubsection{Longitudinal displacement}

The mutual distance between adjacent particles in the trapping state is significantly larger than the cut-off distance.
Therefore, the steric interactions between membrane particles vanish, and only the elastic and dipolar interactions are relevant. 

Assuming that $\kappa \ne 1$, 
Eq.~\eqref{discreteDisplacementsU} that governs the final steady-state membrane displacement along the $x$~direction, for $1 < i < M$, can be written as
\begin{equation}
	 u_{i+1}-2u_i+u_{i-1} =  0 \, . \label{discreteDisplacementsU2}
\end{equation}
The latter expression is subject to the boundary condition $u_{M-1}-3u_{M}=Kh$, which follows from setting $i=M$ in Eq.~\eqref{discreteDisplacementsU} and using the fact that $u_{M+1}=-u_{M}$ as required by symmetry considerations.
Here, we have defined for convenience the dimensionless number
\begin{equation}
	K = \frac{P_0 }{4(\kappa-1)(1+\delta)}  \, ,
\end{equation}
where we have used the approximation $\sin\alpha \simeq 1/\left(1+\delta \right)$.
The solution of the resulting linear homogeneous second-order recurrence problem satisfying the zero-displacement boundary condition $u_1=0$, is given by
\begin{equation}
	\frac{u_i}{h} = -\frac{i-1}{N-1} \, K \, .
\end{equation}
The maximum displacement occurs for $i=M$ and amounts to $u_{M} = -(M-1)Kh/(2M-1)$.

For $\kappa=1$, the dipolar forces are balanced by the elastic forces.
Consequently, the membrane to linear order primarily undergoes motion along the transverse direction.

We further note that for $\kappa \leq 1$ the elastic forces cannot stabilize the system as the dipolar attraction overwhelms the elastic repulsion.
Since its ends are fixed, the membrane would tear itself apart.
The steric repulsions in this situation prevent the collapse of the system.

\subsubsection{Transverse displacement and dipole orientation}

We next consider the displacement field induced along the transverse direction and examine the rotation of the dipoles.
For $1 < i < M$, Eqs.~\eqref{discreteDisplacementsRho} and \eqref{discreteOrientation} are written in the steady trapping state as
\begin{subequations}\label{steadyDiscreteOrientation}
	\begin{align}
	\frac{1}{h} \left( \rho_{i+1}-2\rho_i+\rho_{i-1} \right) - \frac{1}{4} \left( \phi_{i+1}-\phi_{i-1}\right) &=0\, , \\
	\frac{1}{h} \left( \rho_{i+1}-\rho_{i-1} \right) - \frac{1}{3} \left( \phi_{i+1}+4\phi_i+\phi_{i-1} \right) &= 0 \, .
	\end{align}
\end{subequations}

For the solution of the coupled recurrence relations at hand, it is convenient to rearrange the equations in such a way as to decouple the transverse displacement from the dipole orientation.
To that end, we define the displacement gradient as $D_i = (\rho_i-\rho_{i-1})/h$.
Accordingly, Eqs.~\eqref{steadyDiscreteOrientation} can be rewritten as
\begin{subequations}\label{steadyDiscreteOrientationYuri}
	\begin{align}
	D_{i+1}-D_i &= \frac{1}{4} \left( \phi_{i+1}-\phi_{i-1} \right) \, , \\
	D_{i+1}+D_i &= \frac{1}{3} \left( \phi_{i+1}+4\phi_i+\phi_{i-1} \right) \, .
	\end{align}
\end{subequations}
Then, Eqs.~\eqref{steadyDiscreteOrientationYuri} can be rearranged to obtain
\begin{equation}\label{relationZwischenDundPhi}
	D_i = \frac{2}{3} \, \phi_{i-1} + \frac{7\phi_i+\phi_{i-2}}{24}  
	=
	\frac{2}{3} \, \phi_i + \frac{ 7\phi_{i-1}+\phi_{i+1}}{24}  \, .
\end{equation}

The latter equation can further be rearranged to obtain the following recurrence relation for the orientation field,
\begin{equation}
	\phi_{i+1}-\phi_{i-2} + 9 \left( \phi_i-\phi_{i-1} \right) = 0 \, . 
	\label{EqPhi}
\end{equation}

In order to solve the resulting linear homogeneous third-order recurrence problem and find the general term of $\phi_i$, we use the classical approach based on the \textit{distinct roots theorem}~\cite{rudin76}.
Correspondingly, we search for solutions of the recurrence relation in the form of $\phi_i = c/p^i$.
Substituting into Eq.~\eqref{EqPhi} yields the characteristic equation of the recurrence problem, 
\begin{equation}
	p^3+9p^2-9p-1=0 \, , 
\end{equation}
the solutions of which, often called  the characteristic roots of the recurrence relation, are $p=1$ and $p_\pm :=-5\pm 2 \sqrt{6}$.
Then, the general solution for the orientation field is given by
\begin{equation}
	\phi_i = C + C_- p_-^i + C_+ p_+^i \, , \label{phiLoesung}
\end{equation}
where the constants $C_\pm$ and $C$ are to be determined from the boundary conditions.
We note that $p_+$ and $p_-$ are the multiplicative inverse of each other, i.e., $p_+ p_- = 1$.

Upon substitution of the expression of the orientation field given by Eq.~\eqref{phiLoesung} into Eq.~\eqref{relationZwischenDundPhi}, the general solution for the displacement gradient is obtained as
\begin{equation}
	D_i = C + 
		C_- \left( -1+\tfrac{\sqrt{6}}{2} \right) p_-^i 
		+ 
		C_+ \left( -1-\tfrac{\sqrt{6}}{2} \right) p_+^i \, . \label{DLoesung}
\end{equation}

For the determination of the three unknown coefficients $C$~and $C_\pm$, we make use of the boundary conditions,
\begin{subequations}\label{BCs2}
	\begin{align}
		3D_2 - \left( \phi_2+2\phi_1 \right) &= 0 \, , \\
		D_M - \frac{1}{4} \left( \phi_{M}+\phi_{M-1} \right) &= \frac{P_0}{2} \, , \\
		D_M - \phi_{M} - \frac{\phi_{M-1}}{3} &= 0 \, ,
	\end{align}
\end{subequations}
after noting that $\rho_{M+1}=\rho_{M}$ and $\phi_{M+1}=-\phi_{M}$.
Here, $P_0=\mu F_0 h/A$ is the dimensionless parameter defined earlier in Eq.~\eqref{P0Def}.

Next, from Eqs.~\eqref{phiLoesung} through \eqref{BCs2}, the unknown coefficients are determined as
\begin{align}
	C &= -W \left(12 Q_{M-1} + 117 Q_M + \sqrt{6} \left(5 S_{M-1} + 48 S_M \right)\right)  \, , \notag \\
	C_\pm &= W \left(\pm 12+5\sqrt{6} \right) \, , \notag 
\end{align}
where we have defined
\begin{equation}
	S_i = p_+^i + p_-^i \, , \qquad
	Q_i = p_+^i - p_-^i \, .
\end{equation}
Moreover, $W=P_0/ \left(3Q_M+\sqrt{6}S_M\right)$.

The transverse displacement field of the $i$th membrane particle can then be calculated from the displacement gradient as
\begin{equation}
	\rho_i = h \sum_{j=2}^{j=i} D_j \, , 
\end{equation}
which, using $\rho_1=0$, reads
\begin{align}
		\frac{\rho_i}{h} &= (i-1)C + 
			C_- \left( -1+\tfrac{5\sqrt{6}}{12} \right)
			\left( 49+20\sqrt{6}-p_-^{i+1} \right) \notag \\
			&+ 
			C_+ \left( 1+\tfrac{5\sqrt{6}}{12} \right)
			\left( -49+20\sqrt{6}+p_+^{i+1} \right)  \, . \label{rhoLoesung}
\end{align}

In the limit of $M\to\infty$ (and thus $h \to 0$ for fixed~$L$), we get $C = P_0$ and $C_- = C_+ = 0$.
Defining a continuum variable as $x/L= \left( (i-1)/(M-1) - 1 \right)/2 $ for $1\le i \le M$ such that $x/L \in [ -1/2, 0 )$, Eq.~\eqref{rhoLoesung} can be written in the continuum limit, for~$x$ notably smaller than zero, as
\begin{subequations}\label{solutionDiskret}
	\begin{align}
		\lim_{M\to\infty} \phi (x) &= P_0  \, , \\
		\lim_{M\to\infty} \rho (x) &= P_0 \left( \frac{L}{2} + x \right) \, .
	\end{align}
\end{subequations}

It is worth mentioning that our approximation is valid in the small deformation regime for which  $P_0 \ll 1$.
From parity considerations, it follows that $\phi(-x) = -\phi(x)$ and $\rho(-x)=\rho(x)$.
Thus, the transverse displacement reaches its maximum value at the membrane center, for $x=0$.

In the following, we will approach the problem differently by utilizing a continuum description of the governing equations to yield analytical expressions for the membrane deformation not only in the steady state, but also in the transient state.

\subsection{Continuum description}

In order to obtain a continuum description of the membrane deformation and dipole orientations, we present the transverse displacement field in the form $\rho_{i+s}=\exp(shD)\rho(x)$, and analogously for $u_{i+s}$ and $\phi_{i+s}$, wherein~$s$ is a relative integer, and $D := \partial/\partial x$ denotes the differential operator with respect to the spatial coordinate.
Expanding the exponential argument in powers of $shD$, we obtain for~$\rho_{i+s}$ up to second order~\cite{kumar65}
\begin{equation}
	\rho_{i+s} = \left( 1 + sh \, \frac{\partial}{\partial x} +\frac{(sh)^2}{2} \frac{\partial^2}{\partial x^2} + \dots \right) \rho(x) \, , 
\end{equation}
and analogously expressions for $u_{i+s}$ and $\phi_{i+s}$.

Using this representation, Eqs.~\eqref{discreteDisplacements} can be written in the continuum limit as
\begin{subequations}\label{KontinuumGleischungenKraeftegleichgewicht}
	\begin{align}
		u_{,t} &= 2A \left( \kappa -1 \right) u_{,xx} \, , \label{verdraengungGleischungU} \\
		\rho_{,t} &= A
		\left( \rho_{,xx} - \frac{\phi_{,x}}{2} \right) + \bT \left( F_0 - \frac{\yPt}{\bP} \right) h \,\delta(x) \, ,	\label{verdraengungGleischungRho}
	\end{align}
\end{subequations}
for $-L/2 \le x \le L/2$.
Here, commas in the subscripts denote partial derivative with respect to the arguments listed in the subscripts.
We have neglected the steric interactions along the longitudinal direction as they usually have a vanishing contribution to the force balance in the trapping state, during which the membrane is stretched.
In addition, the discrete force $F_i = F \left( \delta_{i,M} + \delta_{i,M+1} \right)$ has now been transformed into a point force $2Fh \, \delta(x)$ along the $y$ direction, where the prefactor $h$ has been introduced so as to ensure the right physical dimension. 
Accordingly, $\alpha \to 0$ holds in the continuum limit, since $a\to 0$ leads to $\delta \to \infty$ for~$R$ remaining finite.
Thus the longitudinal component of the force $F_\parallel$ vanishes.

Similarly, the continuum version of the equation governing the orientation dynamics of the dipoles, given in a discrete form by Eq.~\eqref{discreteOrientation}, reads
\begin{equation}
	\phi_{,t} = B
	\left( \rho_{,x} - \phi \right) \, .
	\label{orientierungsdynamik}
\end{equation}

Eqs.~\eqref{KontinuumGleischungenKraeftegleichgewicht} and \eqref{orientierungsdynamik} are subject to the initial conditions at $t=0$ of vanishing displacement and orientation, in addition to the boundary conditions of zero displacement and torque at $x = \pm L/2$.
It is worth mentioning that $A$ and $B$ are considered here as constant membrane properties and are therefore not affected by the limit $h \to 0$.

\subsubsection{Steady state}

We first look for analytical solutions of the continuum model equations in the steady state of motion.
It follows from Eq.~\eqref{verdraengungGleischungU} that the steady longitudinal displacement in the trapping state satisfies $u_{,xx} = 0$.	
Since $u(x=0)=0$, as required by symmetry considerations, the longitudinal displacement necessarily vanishes upon application of the boundary conditions.
Therefore, the membrane particles only displace along the $y$ direction in the considered continuum limit.

As for the transverse displacement, Eq.~\eqref{verdraengungGleischungRho} simplifies in the steady state to
\begin{equation}
		\rho_{,xx} - \frac{\phi_{,x}}{2} + P_0 \, \delta(x) = 0 \, , 
\end{equation}
while Eq.~\eqref{orientierungsdynamik} leads to $\phi = \rho_{,x}$.
As a result, the steady orientation of the dipoles is solely given by the displacement gradient.
The present situation is analogous to that known in the context of Kirchhoff--Love theory of elastic beams or plates~\cite{timoshenko59}.
Thus, the transverse displacement of the continuous membrane is governed by the following second-order differential equation,
\begin{equation}
	\rho_{,xx} + 2 P_0 \, \delta(x) = 0 \, ,
\end{equation}
the solution of which (that satisfies the boundary conditions) is given by
\begin{subequations} \label{solutionKontinuum}
	\begin{align}
		\rho(x) &= P_0 \left( \frac{L}{2} - |x| \right) \, , \\
		\phi(x) &= -P_0 \, \sgn (x) \, ,
	\end{align}
\end{subequations}
where $\sgn(x) := x/|x|$ denotes the sign function.
These results are in full agreement with Eqs.~\eqref{solutionDiskret} that have been obtained for $x<0$ by taking the corresponding continuum limit in the discrete description.

The membrane undergoes a maximum deformation at its center, which, for $h=L/N$ and $h/a= 2^{7/6}$, is given by
\begin{equation}
	\frac{\rho_\mathrm{Max}}{L} = \frac{P_0}{2}
	= 4\pi c \, \frac{a^4 F_0}{\mu_0 m^2} \, , \label{maxDisplacementKontinuum}
\end{equation}
where $c= 2^{5/3} /3 \approx 1.06$ is a numerical prefactor.
The latter result indicates that the maximum deflection of the membrane scales linearly with the magnitude of the active force but does not depend on the nature of the steric interactions causing the membrane to deform.

\begin{figure}
\begin{center}
\includegraphics{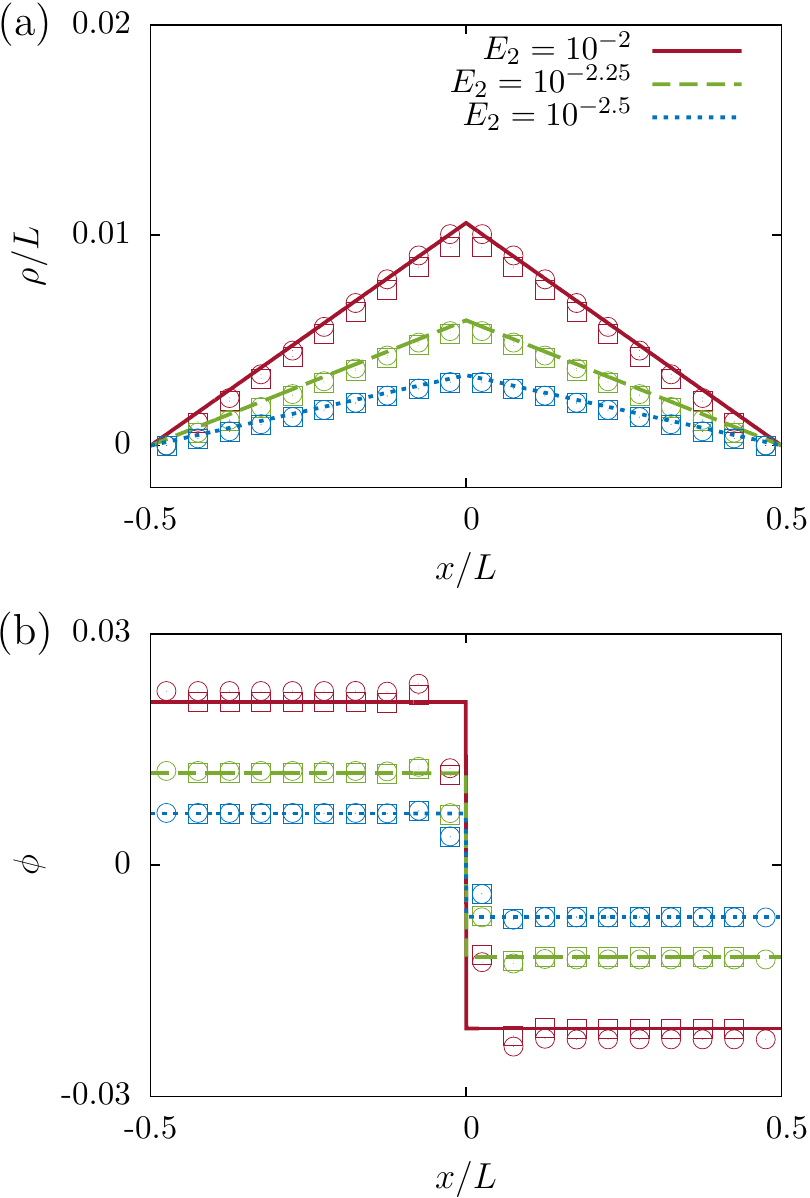}
\end{center}
\caption{(Color online) 
Steady-state solutions in the trapping state.
$(a)$ Scaled membrane deformation $\rho / L$, and 
$(b)$ local membrane orientation $\phi$ as functions of $x$ (the self-driven particle is located at $x=0$), both for systems with $E_1 = 1$, $N = 20$, and varying values of $E_2$.
Circles indicate the results of numerical simulations obtained by solving the nonlinear dynamical equations,
rectangles denote the solutions of the recurrence problem given by Eqs.~\eqref{phiLoesung} and \eqref{rhoLoesung},
and solid lines are the analytical predictions described by Eqs.~\eqref{solutionKontinuum} obtained from a continuum formulation.
All these approaches lead to triangular profiles for $\rho(x)$ and square-like ones for $\phi(x)$, showing strong quantitative agreement without the introduction of any fitting parameters.
}
\label{FigSteady}
\end{figure}

In Fig.~\ref{FigSteady}, we present the steady-state profiles of $(a)$ the transverse displacement $\rho(x)$ and $(b)$ the orientation $\phi(x)$ for various values of $E_2$, while keeping the other parameters constant at $E_1=1$ and $N=20$.
Here, the numerical solutions of the nonlinear equations are indicated by circles, and the results of the corresponding recurrence solution of the linear discrete problem -- closely matching the numerical solution -- are denoted by squares.
Solid lines present the continuum solutions for the same set of parameters.

While the continuum description always leads to ideal triangular and, respectively, square profiles for
$\rho(x)$ and $\phi(x)$, the numerical solution of the nonlinear problem shows deviations from these shapes.
The differences are most probably due to the finite size of the active particle which has not been taken into account in the present continuum description.
Finally, we remark that even though no fitting parameters have been introduced,
the results still closely match each other, reinforcing the applicability of our approximate analytical approach 
to predict the shape of our minimal membrane model under the influence of a localized destroying force.

\subsubsection{Transient behavior}

Having presented analytical solutions of the continuum equations of motion in the steady state, assessed the appropriateness and judged the accuracy of our linearized analytical theory, we next address the membrane deformation and dipole orientation in the transient regime.
The solution to this mathematical problem can be obtained by \emph{finite} Fourier transforms in space of the governing equations, and solving the resulting ordinary differential equations in time.

For this purpose, we define the basis functions
	\begin{equation}
		c_q (x) = \cos \left( H_q x \right) \, , \qquad 
		s_q (x) = \sin \left( H_q x \right) 
		\, , 
	\end{equation}
where $H_q = (2q-1)\pi/L$ with $q = 1, 2, \dots$ denoting the variable that sets the coordinates in Fourier space.
Then the displacement and orientation fields can be expressed in terms of Fourier series in space as~\cite{bracewell99}
\begin{subequations}\label{phiRhoLoesungTransient}
	\begin{align}
		\rho(x,t) &= \frac{2}{L} \sumOneQ \hrho(q,t) \, c_q (x) \, , \\
		\phi(x,t) &= \frac{2}{L} \sumOneQ \hphi(q,t) \, s_q (x) \, , 
	\end{align}
\end{subequations}
where $\hrho$ and $\hphi$ are the Fourier coefficients, defined as
\begin{subequations}
	\begin{align}
		\hrho (q,t) &= \int_{-\frac{L}{2}}^{\frac{L}{2}} \rho(x,t) \, c_q (x) \, \Intd x \, , \\
		\hphi (q,t) &= \int_{-\frac{L}{2}}^{\frac{L}{2}} \phi(x,t) \, s_q (x) \, \Intd x \, .
	\end{align}
\end{subequations}
The form of the Fourier representation given by Eqs.~\eqref{phiRhoLoesungTransient} follows from the boundary conditions to ensure at any time that $\rho (\pm L/2,t) = 0$ and $\phi_{,x}(\pm L/2, t)=0$.
We note that the basis functions $c_q(x)$ and $s_q(x)$ satisfy the orthogonality relations
\begin{equation}
	\int_{-\frac{L}{2}}^{\frac{L}{2}} c_p(x) c_q(x) \, \Intd x 
	=\int_{-\frac{L}{2}}^{\frac{L}{2}} s_p(x) s_q(x) \,\Intd x 
	= \frac{L}{2} \, \delta_{pq} .
\end{equation}

Transforming Eqs.~\eqref{verdraengungGleischungRho} and \eqref{orientierungsdynamik} into spatial Fourier space yields 
\begin{subequations}\label{diffRhoT_diffPhiT}
	\begin{align}
		\frac{\hrho_{,t}}{A} &= -H_q \left(  H_q \hrho + \frac{\hphi}{2} \right)
		+ P_0 \left( 1 - \frac{\yPt}{v_0}  \right)
		 \, , \label{diffRhoT_Fourier} \\
		\frac{\hphi_{,t}}{B} &= -H_q  \hrho - \hphi \, ,  
	\end{align}
\end{subequations}
where $v_0 = \bP F_0$ is the bulk velocity of the active particle.

For a closure of the above set of equations, we require that the instantaneous distance between the self-driven particle and the membrane center remains constant during the system evolution, such that ${\yP}_{,t} = \rho_{,t}(x=0,t) $.
However, in order to be able to make analytical progress, we further assume that after a brief transient evolution, $|\yPt| \ll v_0$ holds, and thus the term involving~$\yPt$ can be neglected.
This is equivalent to assuming that the active particle instantaneously attains its terminal velocity when the interaction with the membrane takes place.

The solution of the system of differential equations given by Eqs.~\eqref{diffRhoT_diffPhiT} can more easily be obtained using the Laplace transform technique~\cite{widder15}.
In the following, the Laplace-transformed function pairs are distinguished only by their argument while the hat is reserved to denote the spatial Fourier transforms.
By employing the initial conditions $\hrho(q,t=0)=\hphi(q,t=0)=0$, we obtain
\begin{subequations}
	\begin{align}
		\frac{s}{A} \, \hrho(q,s) &= -H_q \left( H_q \hrho(q,s) + \frac{\hphi (q,s)}{2} \right)
				+ \frac{P_0}{s} \, , \\
		\frac{s}{B} \, \hphi(q,s) &= -H_q \hrho(q,s) - \hphi(q,s) \, .	
	\end{align}
\end{subequations}
Solving these equations for $\hrho (q,s)$ and $\hphi (q,s)$ yields
\begin{subequations}
	\begin{align}
		\hrho (q, s) &= \frac{2A(B+s) P_0}{Q} \, , \\
		\hphi (q, s) &= -\frac{2H_qAB P_0}{Q} \, , 
	\end{align}
\end{subequations}
where the denominator is given by
\begin{equation}
	Q = s \big( 2s^2 + 2(B+A H_q^2)s + AB H_q^2 \big) \, . \notag
\end{equation}

The inverse Laplace transform can readily be obtained from the standard approach of partial fraction decomposition and using tables of Laplace transforms, which yields
\begin{align}
	\hrho (q,t) &= \frac{2P_0}{H_q^2} \Bigg( 
	1-e^{-\beta t}
	\left( \cosh \left( \tau t \right) 
	+ \frac{B}{2\tau} \, \sinh \left( \tau t\right) 
 \right)
	\Bigg) \, , \notag \\
	\hphi (q,t) &= -\frac{2 P_0}{H_q} \Bigg(
	1- e^{-\beta t } 
	\left( \cosh(\tau t) + \frac{\beta}{\tau} \, \sinh (\tau t) \right)
	\Bigg) \, , \notag
\end{align}
where we have defined the parameters~$\tau$ and $\beta$, with inverse time dimension, as
\begin{equation}
	\tau = \frac{1}{2} \sqrt{B^2+A^2 H_q^4} \, , \qquad
	\beta = \frac{1}{2} \left( B+AH_q^2 \right) \, .
\end{equation}

\begin{figure}
\begin{center}
\includegraphics{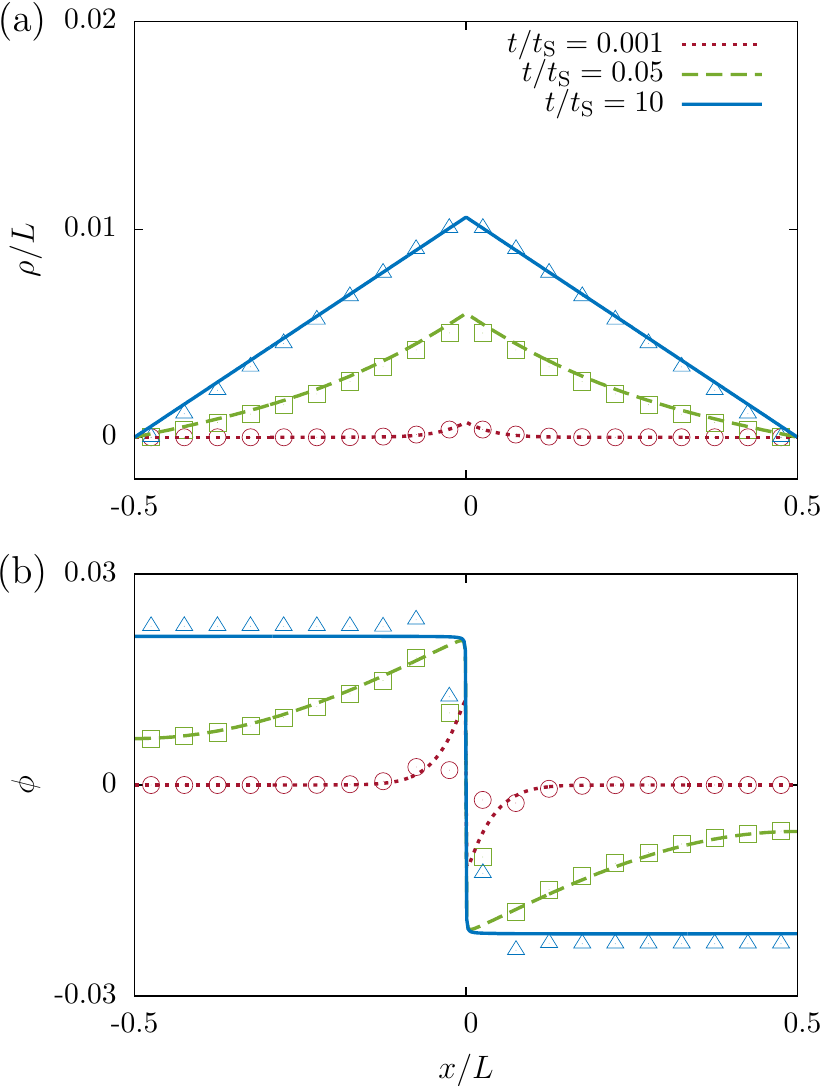}
\end{center}
\caption{(Color online) 
Dynamic solutions for the trapping scenario.
$(a)$ Scaled transient membrane deformation profile $\rho (x,t) / L$ and $(b)$ membrane orientation profile $\phi(x,t)$ calculated at times $t/t_\mathrm{S}=0.001,0.05,10$, where $t_\mathrm{S}=\eta L^3 / \epsilon$ for $N=20$, $E_1=1$, and $E_2=10^{-2}$.
Here, symbols are numerical simulation results, and solid lines give the corresponding analytical results given by Eqs.~\eqref{phiRhoLoesungTransient}.
Both approaches show qualitative and quantitative agreement in their description of the transition from the small central perturbations at early times to the steady state for $t \to \infty$ (also see Fig.~\ref{FigSteady} for a detailed display of the latter).
}
\label{FigTransient}
\end{figure}


A typical transient behavior is shown in Fig.~\ref{FigTransient} presenting $(a)$ the membrane transverse displacement $\rho(x,t)$ and $(b)$ the dipole orientation $\phi(x,t)$ at various times $t$ using the parameters $E_1=1$, $E_2=10^{-2}$, and $N=20$.
Here, symbols indicate the numerical solutions for a discrete membrane, and
solid lines represent the analytical solutions of the continuum theory outlined above.
Again, without fitting parameters, there is strong qualitative and quantitative agreement between both approaches.

The transverse displacement profile features at early times a small central dent, 
which then more and more expands as time evolves.
This leads to a significant kink at the center and, finally, to the triangular shape in the steady state.
At all times, the symmetry $\rho(-x,t)=\rho(x,t)$ is fulfilled.
Similarly, for the dipole orientation, smooth transitions take place from a small ``orientation jump'' in the center and vanishing initial orientations elsewhere to a full-chain square-like profile in the steady state.
The discrete case features a significantly less pronounced change in orientation for the two central spheres at all times.

Finally, we address the transient behavior in the particular situation of fast orientational relaxation, for which $B \gg A q^2$. 
Setting $\phi_{,t}=0$ in Eq.~\eqref{orientierungsdynamik} yields
\begin{equation}
	\phi = \rho_{,x} \, . \label{fast_orientational_relaxation}
\end{equation}
Accordingly, the dipole orientation follows instantaneously the slope of the membrane.
Inserting Eq.~\eqref{fast_orientational_relaxation} into Eq.~\eqref{verdraengungGleischungRho} yields
\begin{equation}
	\rho_{,t}= \frac{A}{2} \, \rho_{,xx} + \mu h F_0 \, \delta(x) \, ,
\label{heat}
\end{equation}
where $\yPt$ has been neglected along the same lines as above. 

Eq.~\eqref{heat} has the form of a diffusion equation with a point source localized in space, subject to the initial condition $\rho(x,t=0)=0$, in addition to the Dirichlet-type boundary conditions $\rho(x=\pm L/2,t)=0$.
The solution of this equation has been obtained by Sommerfeld~\cite{sommerfeld49} and is expressed as
\begin{equation}
	\rho(x,t) = \frac{A P_0}{2 L} \int\limits_{0}^t \Bigg( \vartheta\left(\frac{x\vphantom{L}}{2L},t'\right)-\vartheta\left(\frac{x+L}{2L},t'\right) \Bigg) \, \Intd t' \, , 
\end{equation}
with Jacobi theta functions~\cite{abramowitz72}
\begin{equation}
	\vartheta(\xi,t) = 1 + 2 \sum\limits_{n=1}^\infty  e^{- \delta_n t} \cos\left( 2 n \pi \xi\right) \, ,
\end{equation}
where we have defined 
\begin{equation}
	\delta_n = \frac{n^2 \pi^2 A}{2L^2} \, .
\end{equation}

This leads to the scaled displacement
\begin{equation}
	\frac{\rho(x,t)}{L} = \frac{4P_0}{\pi^2} \sum\limits_{n=1}^\infty \frac{1 - e^{- \delta_{2n-1} t}}{(2n-1)^2} \, \cos\left( (2n-1) \frac{\pi x}{L}\right) \, ,
\end{equation}
which reproduces the steady-state solution given by Eq.~\eqref{rhoLoesung} as shown below.
In particular, the long-time behavior is dominated by the first term ($n=1$), which approaches the limit exponentially with a characteristic decay time  $2L^2/(\pi^2 A)$. 
Additionally, the orientation follows forthwith from Eq.~\eqref{fast_orientational_relaxation} as
\begin{equation}
	\phi(x,t) = - \frac{4P_0}{\pi} \sum\limits_{n=1}^\infty \frac{1 - e^{- \delta_{2n-1} t}}{2n-1} \, \sin\left( (2n-1) \frac{\pi x}{L}\right) \, .
\end{equation}

In the limit $t\to\infty$, we obtain
\begin{subequations}
	\begin{align}
		\lim_{t\to\infty} \frac{\rho(x,t)}{L} &=
		\frac{4P_0}{\pi^2} \sum_{n=1}^{\infty} \frac{\cos\left( (2n-1) \frac{\pi x}{L}\right)}{(2n-1)^2} \, , \\
		\lim_{t\to\infty} \phi(x,t) &= -\frac{4P_0}{\pi} \sum_{n=1}^{\infty} 
		\frac{\sin\left( (2n-1) \frac{\pi x}{L}\right)}{2n-1} \, , 
	\end{align}
\end{subequations}
which correspond, respectively, to the Fourier series representation of the triangle and square waves functions of frequency $2\pi/L$.
The maximum membrane displacement is calculated as
\begin{equation}
	\lim_{t\to\infty} \frac{\rho(0,t)}{L} = \frac{4P_0}{\pi^2} \sum_{n=1}^{\infty}  \frac{1}{(2n-1)^2}
	= \frac{P_0}{2} \, , 
\end{equation}
in agreement with the result obtained earlier from the steady differential equations, as given by Eq.~\eqref{maxDisplacementKontinuum}.

\section{Conclusions}
\label{sec:conclusions}

In this article we explored the interactions between an active particle and a minimal model membrane.
Since we concentrate on a two-dimensional setup, our results could, for instance, in experiments be readily compared with the behavior of a self-driven particle on a substrate and colliding with a straightened chain of mutually attractive dipolar spheres. 
We demonstrated that the particle may either get trapped by the membrane or penetrate through it, where the membrane can either be permanently damaged or recover by self-healing.
State diagrams are presented that carefully map out which state occurs as a function of only a few generic parameters; membrane elasticity, bending stiffness, strength and size of the active particle.
Our analytical theory further predicts the shape and the dynamics of the membrane, in close quantitative agreement with our numerical simulations.
Our results suggest that the microscopic details of the interactions among membrane components (particles) are largely insignificant to the overall behavior of the membrane. 
Thus, our results might be broadly applicable to describe experiments of micro-swimmers interacting with membranes, such as synthetic microbots colliding with a lipid bilayer, or microbes with a membrane synthesized from dipolar microparticles. 
In this context, it would be interesting to extend our model to account for Brownian noise acting on the membrane and the self-driven particle as well. 
This might, for instance, support the membrane in healing after being destroyed by the penetrating active particle.

\begin{acknowledgments}
ADMI, AMM, and HL thank Joachim Clement for a stimulating discussion.
The authors are indebted to Maciej Lisicki and Shang Yik Reigh for helpful comments and suggestions, and gratefully acknowledge support from the DFG (Deutsche Forschungsgemeinschaft) through the projects DA~2107/1-1, SCHO~1700/1-1, ME~3571/2-2, and LO~418/16-3.
SG gratefully acknowledges funding from the Alexander von Humboldt Foundation.
The work of AJTMM was supported by a cross-disciplinary fellowship from the Human Frontier Science Program Organization (HFSPO - LT001670/2017).
FGL acknowledges support from the Millennium Nucleus \enquote{Physics of Active Matter} of the Millennium Scientific Initiative of the Ministry of Economy, Development and Tourism, Chile.
\end{acknowledgments}

%

\end{document}